\documentclass{article}

\usepackage{arxiv}

\usepackage{graphicx}
\usepackage[utf8]{inputenc} 
\usepackage[T1]{fontenc}    
\usepackage[colorlinks,allcolors=blue]{hyperref}       
\usepackage{url}            
\usepackage{booktabs}       
\usepackage{amsfonts}       
\usepackage{nicefrac}       
\usepackage{microtype}      
\usepackage{lipsum}		
\usepackage{graphicx}
\usepackage{natbib}
\usepackage{doi}
\usepackage{tabularx,booktabs}
\usepackage{algorithm}
\usepackage{algpseudocode} 
\usepackage{multirow}
\usepackage{xcolor}
\usepackage{multicol}
\usepackage{subcaption}
\usepackage{tikz} 
\usepackage{amsmath}
\usepackage{textcomp}
\usepackage{libertine}
\usepackage{comment}
\usepackage{amssymb}

\algnewcommand\algorithmicforeach{\textbf{for each}}
\algnewcommand\algorithmicdoparallel{\textbf{do in parallel}}
\algdef{S}[FOR]{ForEach}[1]{\algorithmicforeach\ #1\ \algorithmicdo}
\algdef{S}[FOR]{ForEachParallel}[1]{\algorithmicforeach\ #1\ \algorithmicdoparallel}
\algnewcommand{\sIf}[2]{
  \State \algorithmicif\ #1\ \algorithmicthen\ #2}

\graphicspath{graphics}
\usepackage{xspace}

\title{Utilising a Quantum Hybrid Solver for Bi-objective Quadratic Assignment Problems\thanks{Please cite: \protect\doi{10.1145/3638530.3664097}}}

\author{ \href{https://orcid.org/0000-0003-0854-4777}{\includegraphics[scale=0.06]{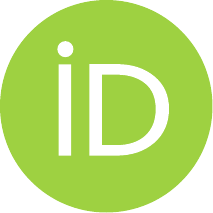}\hspace{1mm}Mayowa Ayodele}\\
	D-Wave Quantum Inc.\\
	Palo Alto\\
	California, USA\\
	\texttt{mayodele@dwavesys.com} \\
}
\date{May 27, 2024}

\renewcommand{\shorttitle}{Utilising a Quantum Hybrid Solver for Bi-objective Quadratic Assignment Problems}

\begin{document}
\maketitle

\begin{abstract}
The intersection between quantum computing and optimisation has been an area of interest in recent years. There have been numerous studies exploring the application of quantum and quantum-hybrid solvers to various optimisation problems. This work explores scalarisation methods within the context of solving the bi-objective quadratic assignment problem using a quantum-hybrid solver. We show results that are consistent with previous research on a different Ising machine.
\end{abstract}



\section{Background}
There has been growing interest in using models that are more flexible than quadratic unconstrained binary optimisation/Ising models often used by Ising machines. One such model is the constrained quadratic model (CQM)\footnote{\url{https://www.dwavesys.com/media/rldh2ghw/14-1055a-a_hybrid_solver_for_constrained_quadratic_models.pdf}}. D-Wave's quantum-hybrid architecture includes a CQM solver which is used in this study. 

In this work, we present a CQM formulation for the bi-objective QAP and compare multiple methods of deriving scalarisation weights. It is common to convert optimisation problems with multiple objectives to single-objective problems when applying Ising machines to solve them. This is often done through a weighted aggregation of the objectives. Setting this weight is not a trivial process. Although, in certain situations, domain experts may be able to suggest weights that could lead to regions of interest in the Pareto front. There has, however, been research on doing this more systematically. A simple approach will be to apply a set of evenly distributed weights to the problem. Previous research has applied the algorithm that supports an Ising machine known as digital annealer to the bi-objective QAP \citep{ayodele2022multi}. Scalarisation technique based on evenly distributed weights was used in \citep{ayodele2022multi}. However, in \citep{ayodele2022study}, a method that attempts to fill the largest gap in the Pareto front was shown to perform better than evenly distributed weights when applying an Ising machine to portfolio optimisation problem instances.  In \citep{ayodele2023applying}, methods that fill the largest gap in the Pareto front were also shown to perform better than evenly distributed weights when applying an Ising machine to unconstrained binary quadratic programming (UBQP) instances.

This work determines whether the conclusions reached in \citep{ayodele2023applying} remain valid for a different solver and a different multi-objective problem. Given that the field of quantum optimisation is rapidly growing while many real-world problems consist of multiple and often conflicting objectives, this work contributes to improving the applicability of quantum and quantum-hybrid optimisation solutions. It should however be noted that the aim of this work is not to compare the CQM solver with other solvers but to compare different scalarisation methods within the context of a quantum-hybrid solver. The rest of this paper contains a formal description of the bi-objective QAP, scalarisation approaches, experimental settings as well as results and conclusions.

\section{Bi-Objective Quadratic Assignment Problem}
\label{sec:prob}
The bi-objective QAP is an extension of the classic QAP that incorporates two distinct objectives. The QAP involves assigning a set of facilities to a set of locations in a way that minimises the sum of the products between flows and distances. In the bi-objective variant, there are two such cost functions to minimise. 
An instance of the bi-objective QAP consists of a $2 \times n \times n$ flow matrix $H=[h_{kij}]$ and a $n \times n$ distance matrix $D=[d_{kl}]$. The two-way one-hot encoding \citep{ayodele2022multi} is used in this study. The cost and constraint functions are defined as

\begin{align}
 c(x) = \sum_{a=1}^{2} \lambda_a \cdot \left ( \sum_{i=1}^{n}\sum_{j=1}^{n}\sum_{k=1}^{n}\sum_{l=1}^{n}h_{k,i,j}d_{k,l}x_{i,j}x_{j,l} \right ) \label{eq:bf}\\
 g_{1,i}(x) = \sum_{j=1}^{n} x_{i,j}  \equiv 1\  \forall\  i \in \left \{ 1,\cdots ,n \right \} \label{eq:g1} \\
 g_{2,j}(x) = \sum_{i=1}^{n} x_{i,j}  \equiv 1\  \forall\  j \in \left \{ 1,\cdots ,n \right \} \label{eq:g2} 
\end{align}

$c(x)$ represents an aggregate objective function of solution $x$. Binary decision variable $x_{i,j}$ is 1 when facility $j$ is placed in location $i$. The scalarisation weights $\lambda_1 + \lambda_2 = 1$.
Constraints $g_{1,i}(x)\ \forall\  i \in \left \{ 1,\cdots ,n \right \}$ and $g_{2,j}(x)\ \forall\  j \in \left \{ 1,\cdots ,n \right \}$ (Eqs. \eqref{eq:g1} and \eqref{eq:g2}) ensures that each row and column of a permutation matrix sums to 1. This ensures that the permutation matrix can be decoded back to a valid permutation list. Constraint $g_{1,i}(x)$ was set as \textbf{discrete}\footnote{\url{https://docs.ocean.dwavesys.com/en/stable/docs_dimod/reference/generated/dimod.ConstrainedQuadraticModel.add_discrete.html}} in the CQM. It should also be noted that penalty weights are not used in this study because the CQM solver does not require penalty weights for hard constraints.

\section{Scalarisation Methods}
\label{sec:scal}
In this work, we consider three approaches of deriving scalarisation weights (i.e., derive $\lambda_1$ and $\lambda_2$ in Eq. \eqref{eq:bf}); \textit{uniform}, \textit{adaptive-averages} and \textit{adaptive-dichotomic}\footnote{Available from \url{https://github.com/mayoayodele/bQAP}}. The \textit{uniform} method uses weights that are evenly distributed. \textit{Adaptive-dichotomic} and \textit{adaptive-averages} methods first minimise each objective individually, then derive the next objective aggregation weights using the information from completed runs of the algorithm. Note that euclidean distance is used to measure the largest gap in the Pareto front during each iteration of \textit{adaptive-dichotomic} and \textit{adaptive-averages} approaches. More detailed explanation of these methods can be found in \citep{ayodele2023applying}.

\section{Results}
\label{sec:res}
One of the most frequently used performance indicators in multi-objective optimisation is the hypervolume metric.
\textit{Hypervolume} \citep{fonseca2006improved} measures the area in the objective space that is dominated by at least one of the points of a non-dominated set and bounded by a given reference point that must be dominated by all points under comparison. We use pymoo's \citep{pymoo} implementation of hypervolume.

\begin{table}[t]
\centering
\captionsetup{font=small}
\caption{Comparing adaptive and uniform methods of generating scalarisation weights: Mean and standard deviation hypervolume values of the returned non-dominated sets across 20 runs are presented. Larger hypervolume values indicate better performance. The maximum objective values in the Pareto optimal set are used as reference sets for the hypervolume. The best mean hypervolume values are presented in bold. Number of weights: 10, Time limit: 5 seconds per scalarisaton. Statistical significance measure: student t-test with 95\% confidence }
\begin{tabular}{@{}ccccccc@{}}
\toprule
\multirow{2}{*}{Instances} & \multicolumn{2}{c}{Uniform} & \multicolumn{2}{c}{Adaptive-averages} & \multicolumn{2}{c}{Adaptive-dichotomic}\ \\ \cmidrule(l){2-7} 
 & Mean HV& Std HV & Mean HV& Std HV & Mean HV& Std HV \\ \midrule
qapStr.25.p75.1 & 2.236E+15	& 4.105E+12	& \textbf{2.242E+15} & 3.362E+12 &	2.238E+15 & 3.150E+12	\\ qapStr.25.p75.2 & 1.128E+16 & 4.640E+13 & \textbf{1.136E+16} & 1.048E+13 & 1.134E+16 & 1.282E+13 \\\midrule
qapStr25.0.1 & 7.982E+17& 1.201E+15& 7.976E+17 & 1.429E+15& \textbf{7.999E+17} & 5.588E+14 \\
qapStr.25.0.2 & \textbf{1.991E+17} & 3.710E+14 & \textbf{1.988E+17} & 5.801E+14 & 1.986E+17 & 6.685E+14 \\\midrule
qapStr25.n75.1 & 3.609E+16& 1.533E+14& 3.569E+16& 5.795E+14& \textbf{3.624E+16}& 4.248E+13\\
qapStr.25.n75.2 & \textbf{4.603E+17} & 4.140E+14 & \textbf{4.602E+17} & 4.838E+14 & \textbf{4.601E+17} & 4.226E+14
  \\ \bottomrule
\end{tabular}
\label{tb:hv}
\end{table}

Table \ref{tb:hv} shows average and standard deviation hypervolume values when the CQM solver is applied to the bi-objective QAP instances\footnote{Available from \url{https://eden.dei.uc.pt/~paquete/qap/}} using different scalarisation approaches.  

\textit{Adaptive-averages} method presents the best average hypervolume on the instances with positive correlation between their objectives (\textit{qapStr.25.p75.1}, \textit{qapStr.25.p75.2}). \textit{Adaptive-dichotomic} method shows the most competitive performance on the instances with negative correlation between their objectives (\textit{qapStr.25.n75.1}, \textit{qapStr.25.n75.2}). There was no one best method for the instances with no correlation between their objectives (\textit{qapStr.25.0.1}, \textit{qapStr.25.0.2}). Overall, \textit{adaptive-averages}, \textit{adaptive-dichotomic} and \textit{uniform} methods present the best (or not significantly different from best) average hypervolume on 4, 3 and 2 of the 6 instances used in this study respectively. These results are consistent with the conclusions in \citep{ayodele2023applying} where \textit{uniform} method was often worse when compared to \textit{adaptive-dichotomic} or \textit{adaptive-averages} methods on bi-objective UBQP instances. The relative performance of \textit{adaptive-dichotomic} and \textit{adaptive-averages} methods on instances with negative and positive correlation between their objectives is also consistent with the results in~\citep{ayodele2023applying}.   

\begin{figure}
    \centering
    \includegraphics[width=\textwidth]{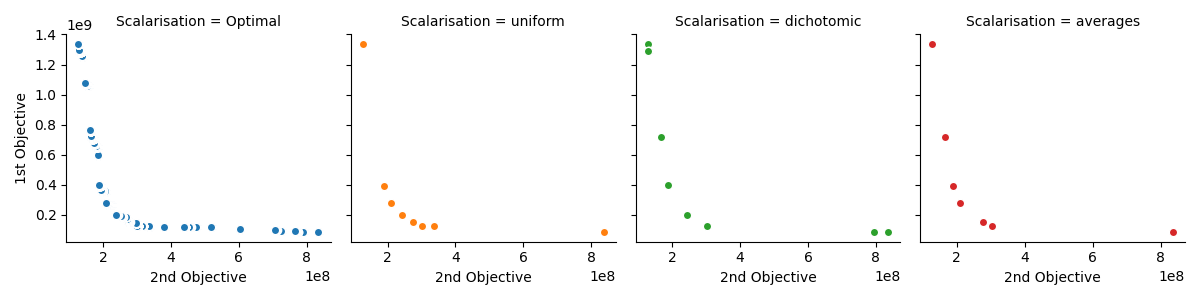}
    \includegraphics[width=\textwidth]{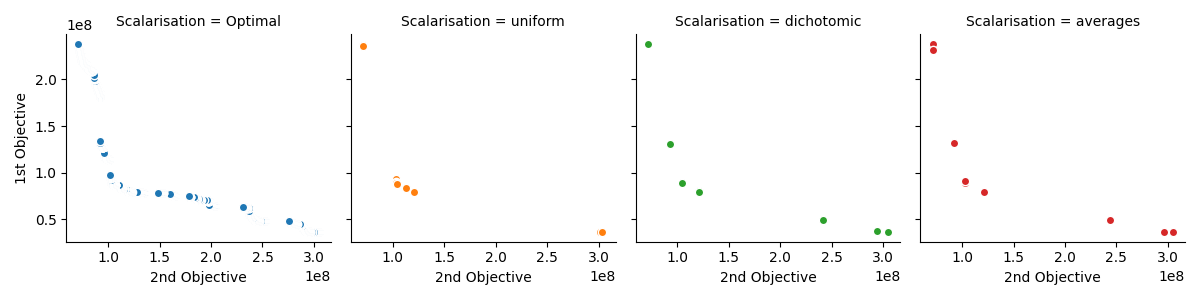}
    \includegraphics[width=\textwidth]{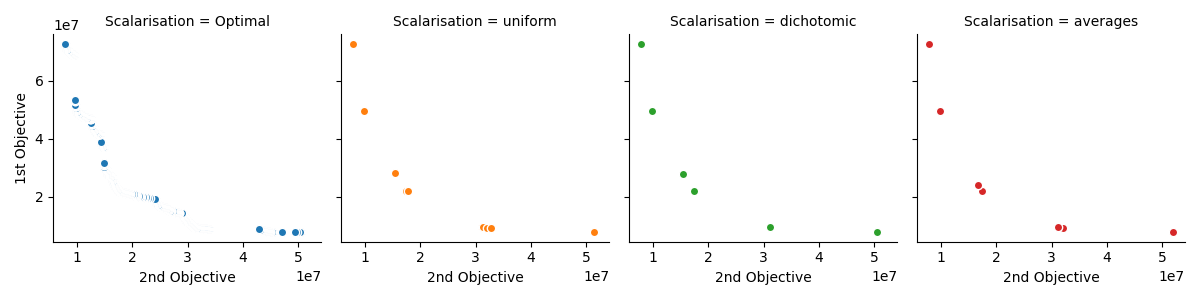}
    \caption{Comparing a single run of all scalarisation methods with the known Pareto Optimal}
    \label{fig:qap}
\end{figure}

For additional context, some results for individual runs are provided in Figure \ref{fig:qap}.


\section{Conclusion}
Three different methods of deriving scalarisation weights have been compared within the context of applying a quantum-hybrid solver to bi-objective QAP instances. Results are consistent with previous research on applying another Ising machine to a different multi-objective problem. Although we have only used a few instances in this study, this work contributes to research on creating a general framework for multi-objective problem solving using Ising machines. For future work, further experiments using more instances of varying sizes will strengthen the conclusions.


\bibliographystyle{unsrtnat}
\bibliography{References}


\end{document}